# Statistic inversion of multi-zone transition probability models for aquifer characterization in alluvial fans

Lin Zhu [1,2], Zhenxue Dai[2], Huili Gong[1], Carl Gable[2], Pietro Teatini[3]

[1]College of Resources Environment and Tourism, Capital Normal University, Laboratory Cultivation Base of Environment Process and Digital Simulation, Beijing 100048, China
[2]Earth and Environmental Sciences Division, Los Alamos National Laboratory, Los Alamos, New Mexico 87545, United States
[3]Deptment of Civil, Environmental and Architectural Engineering, University of Padova, Italy

**Abstract**: Understanding the heterogeneity arising from the complex architecture of sedimentary sequences in alluvial fans is challenging. This paper develops a statistical inverse framework in a multi-zone transition probability approach for characterizing the heterogeneity in alluvial fans. An analytical solution of the transition probability matrix is used to define the statistical relationships among different hydrofacies and their mean lengths, integral scales, and volumetric proportions. A statistical inversion is conducted to identify the multi-zone transition probability models and estimate the optimal statistical parameters using the modified Gauss-Newton-Levenberg-Marquardt method. The Jacobian matrix is computed by the sensitivity equation method, which results in an accurate inverse solution with quantification of parameter uncertainty. We use the Chaobai River alluvial fan in the Beijing Plain, China, as an example for elucidating the methodology of alluvial fan characterization. The alluvial fan is divided into three sediment zones. In each zone, the explicit mathematical formulations of the transition probability models are constructed with optimized different integral scales and volumetric proportions. The hydrofacies distributions in the three zones are simulated sequentially by the multi-zone transition probability-based indicator simulations. The result of this study provides the heterogeneous structure of the alluvial fan for further study of flow and transport simulations.

**Key words**: Multi-zone transition probability; alluvial fan; sediment heterogeneity; structure parameter uncertainty; statistic inversion; indicator simulation

## 1. Introduction

The heterogeneous architecture of aquifers has a great impact on groundwater flow and solute transport in subsurface systems. To characterize aquifer heterogeneity is challenging since in most cases aquifer spatial structures vary greatly while available observation data is limited or not well-distributed in the three-dimensional domain. The sparsely-distributed borehole, outcrop, and geophysical data can only provide partial information to roughly define the aquifer hydrogeologic properties (e.g. the hydrofacies, hydraulic conductivity, and porosity distributions). Therefore, determining the large-scale heterogeneous structures from the above mentioned small-scale data has been recognized as one of the critical unresolved problems in groundwater hydrology (Weissmann and Fogg, 1999; Anderson, 2007; Harp et al, 2008).

Geostatistical and stochastic approaches are common methods to address aquifer characterization and the corresponding uncertainty when borehole or geophysical data do not exist or are sparsely distributed. The Markov chain method provides a conceptually simple and theoretically powerful stochastic model for simulating geological structures with different materials (Carle and Fogg, 1997; Weissmann et al., 1999; Rubin, 2003; Ritzi et al., 2004; Zhang et al., 2006; Dai et al. 2007a; Ye and Khaleel, 2008). A continuous Markov chain is mathematically a transition probability model described by a matrix of exponential functions (Agterberg 1974; Dai et al. 2007b). The Markov chain model has been numerically solved by computing the eigenvalues of the transition rate matrix (Carle and Fogg, 1997). By using the numerical approach of the transition probability models, Proce et al (2004) developed a two-scale Markov chain model to study the heterogeneity of facies assemblages in a regional glacio-fluvial stratigraphy. Sun et al (2008) modeled the sedimentary architecture of the alluvial deposits with a hierarchical structure. However, sometimes there are too many unknown parameters in the transition rate matrix and it is difficult to calculate the eigenvalues and the corresponding spectral component matrices when using the numerical approach (Dai et al. 2007b). Furthermore, without an explicit mathematical formulation, the numerical approach of the Markov chain model cannot reveal the relationships among different hydrofacies and their mean lengths and volumetric proportions. By using a couple of specific assumptions, Dai et al. (2007b) derived an analytical solution of the transition probability matrix, which was successfully applied to aquifer characterization and upscaling transport parameters (Ritzi and Allen-King 2007, Harp et al, 2008; Dai et al., 2009; Deng et al., 2010; 2013; Soltanian et al., 2014; 2015 in press).

This paper develops a statistical inverse framework of a multi-zone transition probability approach for characterizing sedimentary heterogeneity in alluvial fans. Alluvial fans, generally consisting of stream and flooding deposits, can serve as groundwater supply fields since they have abundant water storage and favorable conditions for receiving recharge from precipitation and river leakage. Alluvial-fan aquifers often exhibit spatial variations due to the complicated depositional and digenetic processes which occurred during long-term fan evolution. Investigation of the heterogeneous structures and construction of the stratification sequences of alluvial fans help to improve our understanding of the original sediment transport processes and current hydrogeological properties in fluvial systems (Ritzi et al., 1995; Zappa et al 2006). Wiessmann and Fogg (1999) used transition probability geostatistics in a sequence stratigraphic framework to quantify the facies distributions in an alluvial fan. The flooding deposits of alluvial fans may present spatial zonation along the original sediment transport direction. The volumetric proportions and length distributions of the hydrofacies vary dramatically from the upstream to downstream. Based on the stratigraphic character, geologists usually divide an alluvial fan into three zones: upper fan, middle fan, and lower fan (Miall, 1997). As Wiessmann and Fogg (1999) stated, the assumption of stationarity for a whole alluvial fan is tenuous. Here, we apply the stationarity assumption in each alluvial fan zone since the sediment properties are similar within each zone. The Chaobai alluvial fan in the Beijing Plain, China, is used as an example for developing the multi-zone transition probability approach to characterize the alluvial fan heterogeneity. The sediments of the Chaobai alluvial fan are classified into four categories or hydrofacies: sub-clay and clay, fine sand, medium-coarse sand, and gravel. Multi-zone transition probability models are developed to simulate the sedimentary heterogeneity with different integral scales, volumetric proportions, and mean lengths of the hydrofacies.

We adopt an analytical solution of the transition probability models which incorporates the geologic information on facies proportions, mean lengths and juxtapositional tendencies into geostatistical simulations (Dai et al., 2007b). During this process, the most important step is the inversion of transition probability models to optimally determine the multi-zone aquifer statistical parameters. The statistic inversion of transition probability models is conducted using the generalized output least squares (OLS) criterion to fit the sample transition probability matrix with the analytical solution. The inverse problem is resolved by a modified Gauss-Newton-Levenberg-Marquardt method (Clifton and Neuman, 1982; Dai et al. 2012). The sensitivity equation method is derived to compute the Jacobian matrix for iteratively solving the gradient-based optimization problem (Samper and Neuman 1986; Carrera and Neuman 1986; Dai and Samper, 2004; Samper et al., 2006). Based on the estimated statistical parameters for the multi-zone transition probability models, the three dimensional hydrofacies distributions in different zones are modeled by the transition probability-based indicator simulations.

This study will provide the heterogonous structure of the alluvial fan for the further study of flow and transport simulations.

## 2. Study area

The study area (Figure 1) is located in the upper and middle Chaobai River alluvial fan formed in Late Pleistocene in the Beijing Plain, which covers an area of 1,155 $Km^2$. The mean elevation is about 40 meters above the sea level. The ground surface slope is about 3 ‰ towards the southeast. The Chaobai River is one of the major rivers located at the central axis of the fan, flowing through the Miyun Reservoir at north of the study area. The average precipitation is 624 mm per year (data from 1959 to 2010). About 80% the rainfall concentrates in the period from July to September.

In this study area, the alluvial fan deposits are mainly coarse-grained sediments in the north and gradually change to relatively fine-grained sediments in the south. The Chaobai River alluvial fan is divided into three zones: the upper fan zone, middle fan zone, and lower fan zone. The upper fan zone (also called Zone 1) mainly consists of coarser sands and gravels with a wide range of sediment sizes. Because of the high conductivity in the sand and gravel, the hydrodynamic conditions in the upper fan zone are favorable for receiving recharge from precipitation and river leakage. In the middle fan zone, the strata structure consists of alternate layers of gravels, sands, fine sands, sub-clay and clay. The hydrodynamic inter-connections between different layers are relatively weak. Generally, the cumulative thicknesses of the strata layers and the compressible sediments (sub-clay and clay) in the middle fan zone are greater than those in the upper fan zone. In the western part of the middle fan zone, the total thickness of sediments is generally more than 400 m (Zhu et al. 2013). Since the middle fan zone has a large variation in sediment content and thickness, we divide it into two sub-zones: middle-upper zone and middle-lower zone (or Zone 2 and Zone 3). The lower fan zone mainly consists of the compressible fine sediments (e.g., clay, sub-clay, loam clay, and silt clay). With a very low porosity and hydraulic conductivity, this zone is insignificant for water supply and is not included in this study (Zhu at al., 2015).

The statistics of hydrofacies properties are listed in Table 1 for the above defined three zones (upper fan zone, middle-upper fan zone, and middle-lower fan zone). There are 52 boreholes in Zone 1, 363 boreholes in Zone 2, and 368 boreholes in Zone 3 for characterizing the sedimentary structures. Among all of these boreholes, the maximum exposure depth is about 400 m and the smallest distance between boreholes is about 160 m. Four hydrofacies (e.g., sub-clay and clay, fine sand, medium-coarse sand, and gravel) are classified based on the interpretations of the cores and textural descriptions of the 783 boreholes. In Zone 1 the gravel sediments are dominant with a measured volumetric proportion of 0.56 while the measured proportion of sub-clay and clay deposits in this zone is 0.15. In Zone 2, the proportion of gravel decreases to 0.27 and that of sub-clay and clay increases to 0.42. In Zone 3, the dominant material is sub-clay and clay with a measured proportion of 0.5, and the proportion of gravel decreases to 0.08 (see Figure 2 and Table 1). The maximum cumulative thickness of sub-clay and clay in Zone 1 is much less than that in Zone 3, where the value reaches to 253 m while it is 99 m in Zone 1. The cumulative thickness of gravel in Zone 1 is 336 m and much larger than that in Zone 3. The mean thicknesses of different facies in three zones show the similar features as the maximum cumulative thicknesses (see Table 1).

## 3. Statistic inversion of transition probability models

**Transition probability model**

The transition probability model is a statistical tool for computing probabilities of facies transitions at different lag distances, which incorporates facies spatial correlations, volumetric proportions, juxtapositional tendencies into a spatial continuity model (Agterberg 1974; Carle and Fogg 1996; Ritzi 2000). The transition probability model can improve the implementation of geostatistical simulations of permeability or hydraulic conductivity by taking into account geological and sedimentary information, especially when permeability or hydraulic conductivity measurements are not sufficiently abundant to support the computation of the statistical parameters of permeability or conductivity (Ross 1988; Deutsch and Journel 1992). Transition probability models have been used by geologists and hydrologists to describe the heterogeneity of sedimentary facies for a few decades (e.g. Carle and Fogg 1997; Harp et al., 2008). Recently, Ritzi et al (2004) and Dai et al (2005) incorporated the work of Carle and Fogg (1997) to relate the statistical parameters of the indicator random variables to distributions, geometry, and patterns of the hydrofacies. With two assumptions that the cross-transition probabilities are dictated by facies proportions only and that the juxtapositional tendencies of the facies are symmetric, an analytical solution for the transition probability model was derived by Dai et al. (2007b) with the auto- and cross-transition probability model (Equation 1).

$$t_{ik}(h_\varphi) = p_k + (\delta_{ik} - p_k)\exp(-\frac{h_\varphi}{\lambda\varphi}) \quad (i = 1, 2, \ldots, N; k = 1, 2, \ldots, N), \tag{1}$$

where $t_{ik}(h_\varphi)$ is the transition probability from facies $i$ to facies $k$ in the direction of $\varphi$ with a lag distance $h$, $p_k$ is the volumetric proportion of facies $k$, $\delta_{ik}$ is the Kronecker deltea, $\lambda\varphi$ is the integral scale in the direction of $\varphi$, and $N$ is the number of hydrofacies. Los Alamos National Laboratory developed a geostatistical modeling tool GEOST (Dai et al., 2014) modified from the Geostatistical Software Library (Deutsch and Journel, 1992) and TPROGS (Carle and Fogg, 1997). The geostatistcal tool is employed here to compute sample transition probabilities from the borehole hydrofacies indicator data. The sample transition probabilities will be used for inversion of the multi-zone transition probability models.

**Statistical inversion**

Two types of statistical parameters, volumetric proportion and integral scale, are included in the analytical solutions of the transition probability models (Eq. 1). By estimating these statistical parameters, we are able to fit the computed analytical solutions to the corresponding sample transition probabilities. Let the statistical parameter vector $x = (p_1, p_2, p_3, \ldots, p_N, \lambda_\varphi)$, then, the least-squares criterion $E(x)$ and the corresponding constraints can be expressed as

$$E(x) = Minimize \ \sum_{l=1}^{L}(U_l(x) - F_l)^2, \tag{2}$$

$$\sum_{k=1}^{N} p_k = 1 \ and \ \sum_{k=1}^{N} t_{ik}(h_\varphi) = 1,$$

$$0 \leq p_k \leq 1 \ and \ 0 \leq t_{ik}(h_\varphi) \leq 1 \ (i = 1, 2, \ldots, N; k = 1, 2, \ldots, N),$$

where $L$ is the number of sample transition probability data, $U_l(x)$ is the output of the analytical solution (1), and $F_l$ is the sample transition probabilities. The constraint equations require that the sum of the volumetric proportions of the $N$ facies is one and the sum of the transition probabilities in one row of the transition matrix is also one. The least squares equation (2) can be solved by the modified Gauss-Newton-Levenberg-Marquardt method as

$$X_{j+1} = X_j - (J^T \omega J + \alpha I)^{-1} J^T \omega E(x), \tag{3}$$

where $X_j$ is the vector of parameter values at the $j$-th iteration, $\omega$ is a diagonal weighting matrix, $I$ is the identity matrix and $\alpha$ is the Marquardt parameter, and $J$ is the Jacobian matrix reflecting the sensitivity of the output variable to the statistical parameters. Generally, there are three methods to compute the Jacobian matrix: the finite difference method (Doherty and Hunt, 2009), the variational method (Sun and Yeh, 1990) and the sensitivity equation method (Samper and Neuman, 1989). Here we use the sensitivity equation method to analytically derive the sensitivity coefficients. In Equation 4, $\frac{\partial t}{\partial p}$ and $\frac{\partial t}{\partial \lambda_\varphi}$ are the sensitivity coefficients (or partial derivations) of the transition probability to volumetric proportion and integral scale, respectively.

$$J = \begin{bmatrix} \frac{\partial t_1}{\partial p_1} & \frac{\partial t_1}{\partial p_2} & \cdots & \frac{\partial t_1}{\partial p_N} & \frac{\partial t_1}{\partial \lambda_\varphi} \\ \frac{\partial t_2}{\partial p_1} & \frac{\partial t_2}{\partial p_2} & \cdots & \frac{\partial t_2}{\partial p_N} & \frac{\partial t_2}{\partial \lambda_\varphi} \\ \vdots & \vdots & \ddots & \vdots & \vdots \\ \frac{\partial t_L}{\partial p_1} & \frac{\partial t_L}{\partial p_2} & \cdots & \frac{\partial t_L}{\partial p_N} & \frac{\partial t_L}{\partial \lambda_\varphi} \end{bmatrix}$$

$$\frac{\partial t}{\partial p_k} = 1 - exp\left(-\frac{h_\varphi}{\lambda_\varphi}\right) \quad (k = 1, 2, \ldots, N)$$

$$\frac{\partial t}{\partial \lambda_\varphi} = (\delta_{ik} - p_k) * (-h_\varphi) * exp\left(-\frac{h_\varphi}{\lambda_\varphi}\right) * \lambda_\varphi^{-2} \quad (i = 1, 2, \ldots, N; k = 1, 2, \ldots, N) \quad (4)$$

Prior parameter information is incorporated in the objective function as additional "observation data" and it is also used to define parameter initial values, minimum and maximum bounds, which can help deciding a range of acceptable values that parameters can take during the optimization process. Incorporating prior information into the objective function can also alleviate the ill-posedness and non-uniqueness of inverse problems (Dai and Samper, 2004; Samper at al., 2006). At the beginning of each iteration of the Gauss-Newton-Levenberg-Marquardt method, $\alpha$ is reduced by a factor β in an attempt to push the algorithm closer to the Gauss-Newton method. If this fails to give a reduction in the objective function, $\alpha$ is repeatedly increased by factor β until a reduction is obtained. By comparing parameter changes and objective function improvement achieved in the current iteration with those achieved in the previous iteration, the algorithm can tell whether it is worth undertaking another optimization iteration. If so, the whole process is repeated. Finally, the algorithm stops when a convergent solution is met or the maximum number of iterations is attained.

**Parameter uncertainty analysis**

Analyzing inversion error and quantifying parameter uncertainty are even more important than finding the best-fit parameters because models never exactly fit the data even when the model structures are correct. During the optimization processes, we compute the variances of the estimated parameters, as well as the corresponding covariance matrix $C(p)$,

$$C(p) \cong s^2 (J^T \omega J)^{-1}, \tag{5}$$

where $s^2$ is the total variance of estimated parameters, $C_{ii}$ is the diagonal element of the covariance matrix which represents the variance of parameter $i$, and $\omega$ is the diagonal weighting matrix. Parameter uncertainty is quantified by confidence intervals which is computed from the covariance matrix (5) as expressed by Mishra and Parker (1989) as

$$P_r \left[ |x_i - x_i^*| \leq \sqrt{\chi_\alpha^2} \sqrt{C_{ii}} \right] = 1 - \alpha \tag{6}$$

where $x_i^*$ is the estimated value of the parameter, $x_i$ is the true parameter value, and $\chi_\alpha^2$ is the chi-square statistics corresponding to $M$ degrees of freedom and a confidence interval of $100(1 - \alpha)$ (Carrera and Neuman, 1986).

**Multi-zone indicator simulation models**

Generally, the indicator cross-variogram or indicator covariance is used in geostatistics to determine the spatial variability of indicator variable (Deutsh and Journel, 1992). Here, we incorporate the multi-zone transition probabilities into indicator geostatistical models to simulate the multi-zone architectures of the hydrofacies in the alluvial fan. According the equations derived by Carle and Fogg (1996) and Ritzi et al. (2004), the relationships of the indicator cross-variogram $\gamma_{ik}(h_\varphi)$ and covariance $C_{ik}(h_\varphi)$ with the transition probability can be expressed as:

$$\gamma_{ik}(h_\varphi) = p_i \left\{ \delta_{ik} - \frac{[t_{ik}(h_\varphi) + t_{ik}(-h_\varphi)]}{2} \right\}$$

$$C_{ik}(h_\varphi) = p_i [t_{ik}(h_\varphi) - p_k] \quad (7)$$

Equation (7) indicates that incorporating the transition probability models into the indicator simulations is equivalent to using the indicator cross-variogram or covariance models. The multi-zone transition probability matrices in vertical, dip, and strike directions are computed by the using analytical solutions with the estimated multi-zone statistical parameters from the statistical inversion in the three zones. When the multi-zone indicator simulations are conducted, the corresponding borehole indicator data are used as the conditional data, which means that the simulations honor the known hydrofacies distributions observed in the boreholes.

## 4. Results and Discussion

**Statistic inverse results**

By using the derived statistical inverse methodology, we identified three sets of transition probability models which correspond to the three zones in the alluvial fan. Three sets of statistical parameters are estimated including indicator integral scales for each zone in the vertical, dip, and strike directions, and facies proportions for each hydrofacies. The 95% confidence intervals are also estimated for quantifying the uncertainty of estimated parameters (Table2). The estimated integral scales and facies proportions vary from Zone 1 to Zone 3, which is consistent with the fact that the sedimentary deposits in the alluvial fan are originated by multiple flooding events. The dominant hydrofacies obviously changes spatially and the average grain sizes reduce from Zone 1 to Zone 3 (Figure 2). In Zone 1, the hydrofacies of the gravel and medium-coarse sand are inter-bedded with minor sub-clay and clay facies. The gravel is predominant with an estimated volumetric proportion of 53.3%. The volumetric proportion of the sub-clay and clay is much lower (with an estimated value of 16.7%). This zone has integral scales of about 17 and 618 m in the vertical and dip directions, respectively. This result indicates that with high transport energy in this zone, the average grain sizes, the mean thickness, and length of the sedimentary facies are relatively larger than those in the two down-stream zones. Figure 3 shows the fitting results between the sample and computed transition probabilities in the vertical and dip directions. The vertical analytical transition probabilities computed with the estimated optimal statistical parameters fit the well-defined sample transition probabilities reasonably (Figure 3a).The sample transition probabilities in dip direction are not well defined because the well spacing in that direction is too large, relatively to the mean length of the hydrofacies, to fully capture the variations of the hydrofacies. The plotted sample transition probabilities are sparse with high uncertainty (Figure 3b). Therefore, we used a trial and error method to estimate the integral scale and fix the volumetric proportions to be the same as those estimated from vertical transition probability. Although the fit is not as definitive as in the vertical direction, it is still reasonably matched the volumetric proportions and integral scale in dip direction. The sample transition probability in strike direction is even more sparsely-distributed with many zero values. We do not conduct inverse modeling for the strike transition probability and assume that the integral scale in this direction is a half of the integral scale in the dip direction for all the three zones and the volumetric proportions of the hydrofacies are the same as those in vertical and dip directions.

Zone 2 and Zone 3 both belong to the middle zone of the alluvial fan. But the estimated volumetric proportions of the hydrofacies and the integral scales in these two zones show significant differences. In Zone 2, the multiple strata of the hydrofacies mainly consist of gravel, fine sand, sub-clay and clay. The estimated volumetric proportion of the sub-clay and clay is significant larger than that in Zone 1 and the value increases to 40.9%. The volumetric proportion of fine-grained sand also shows slight increase, however, the volumetric proportion of the gravel decreases sharply to 24% in comparison with that in Zone 1. The integral scales in Zone 2 are 6.2 and 360 m in the vertical and dip directions, respectively, much smaller than those in Zone 1. The reduced grains sizes of the hydrofacies and the increased numbers of the sedimentary layers correspond to the fact that the sedimentary transport energy level decreases from Zone 1 to Zone 2. The transition probability fitting results for the Zone 2 are shown in Figure 4. In Zone 3, the estimated volumetric proportion of the sub-clay and clay increases to 50.2% and that of the fine sand is about 33%, while that of the gravel reduces to only 6.3%. The integral scales in Zone 3 are 5.3 and 319 m in the vertical and dip directions, respectively, much smaller than those in Zones 1 and 2 (Figure 5). Borehole data shows a higher occurrence frequency (or more layers) of the sub-clay and clay facies in Zone 3 than that in Zone 2, which represents the low-energy water-laid sediments.

If an averaged single-zone transition probability model was adopted for the whole alluvial fan, we found that the transition probability cannot describe the sedimentary spatial variations and the statistical properties of the sedimentary features (e.g., the volumetric proportions of the four hydorfacies and the integral scales). The single-zone transition probability model cannot fit the sample transition probabilities well, especially for the facies of gravel (Figure 6). In the averaged single-zone transition probability model, the estimated vertical integral scale is 8.23 m, which is larger than the integral scales in Zone 2 and Zone 3 while much smaller than that in Zone 1. The dominant facies is sub-clay and clay with a volumetric proportion of 42.23%, the fine sand has a proportion of 30.33%, the gravel has a proportion of 18.58%, and the medium-coarse sand has a proportion of 8.86% (Table 3). This volumetric proportion pattern is similar to the average of the sedimentary features in Zone 2 and Zone 3. However, the hydrofacies features in Zone 1 with the dominant gravel material are not reflected. A single-zone transition probability model implies that the hydrofacies distributions in the whole alluvial fan are stationary, which is not true in this example. Therefore, this result demonstrates that an averaged single-zone transition probability model cannot reflect the sedimentary structure variations in the alluvial fan and the multi-zone transition probability models are needed to honor the observed transition relationships of the hydrofacies and to represent the sedimentary architectures in this study area.

**Multi-zone simulations of the alluvial fan structure**

With the identified multi-zone transition probability models we use the code GEOST to simulate the hydrofacies distributions sequentially from Zone 1 to Zone 3. The statistical attributes of the stratigraphy and the borehole indicator data are honored during the indicator simulation processes. The borehole indicator data are described with a vertical interval of 1 m and are used as the hard conditional data. The estimated volumetric proportions and integral scales in the vertical, dip, and strike directions for the three zones are applied as the major statistical parameters for simulating four hydrofacies distributions in the alluvial fan. The simulated hydrofacies distribution in the three dimensional domain is shown in Figure 7.

There is an obvious lithological change from the upper fan zone to middle fan zones. The gravel deposit represented by red color is dominant in the upper fan zone (Figure 7) and it shows good continuity. The sub-clay and clay deposit marked by blue color is discontinuously distributed with a smaller volumetric proportion in this zone.

From the upper fan zone to middle fan zones (Zone 2 and Zone 3), the volumetric proportions of the gravel deposits decrease gradually. The connection of the gravel deposit becomes worse. The simulated multiple layers of the sub-clay and clay, fine and medium-coarse sand deposits are distributed alternately in Zone 2 and Zone 3, which are consistent with what we observed from the boreholes. Multiple aquifers consisted of fine and medium-coarse sands are inter-bedded with sub-clay and clay which has thicknesses around tens of meters and acts as aquitard or confining layers. The increased volumetric proportion of the sub-clay and clay from Zone 1 to Zone 3 decreases the hydraulic inter-connection in vertical, dip and strike directions. The simulated heterogeneous structure will be incorporated into numerical models for next-step study of groundwater flow and transport simulations.

## 5. Summary and Conclusions

Sedimentary architectures in the Chaobai alluvial fan show obvious heterogeneity from the upper fan to lower fan. This paper developed a statistical inverse framework to identify multi-zone transition probability models and use them to characterize and model the spatial heterogeneity. The optimized statistical parameters were estimated with a nonlinear optimization technique. The sensitivity equation method was used to calculate the Jacobian matrix to increase accuracy of the statistical inversion. The uncertainties of the statistical parameters were quantified with 95% confidence intervals.

On the basis of the borehole geological descriptions and log data collected from the alluvial fan, four categories of hydrofacies were identified, including gravel, medium-coarse sand, fine sand, and sub-clay and clay. These hydrofacies vary dramatically in their integral scales and volumetric proportions from the upstream to downstream of the alluvial fan. In the upper fan zone (Zone 1), the dominant hydrofacies is gravel with a volumetric proportion of 53% which is inter-bedded with relatively thin layers of sand and clay. The content of the material of sub-clay and clay is minor with a volumetric proportion of 16%.

In the middle-upper fan zone, the volumetric proportion of the sub-clay and clay increases to 40%, while that of the gravel decreases sharply to 24%. In the middle-lower fan zone, the proportion of gravel decreases further to 6% and that of sub-clay and clay increases to 50%, which is alternatively inter-bedded with multiple thin layers of fine sand and medium-coarse sand. The fine sand and medium-coarse sand have proportions of about 33% and 11%, respectively. The estimated vertical integral scale is the largest in upper fan zone (17 m), which corresponds to the largest mean thickness of the gravel in this zone. The vertical integral scales decrease to 6.2 and 5.3 m, in Zone 2 and Zone 3, respectively.

The averaged single-zone transition probability model does not fit well the sample transition probability, which demonstrates that the multi-zone transition probability models are needed to characterize the complex heterogeneous structures of alluvial fans. The simulated hydrofacies distribution with multi-zone transition probability models represents the sedimentary structures of the alluvial fan reasonably. Since the well spacing is too large to fully catch the facies variations in the dip and strike directions, the estimated integral scales in these two directions contain some degree of uncertainty. Further geological and geophysical work may increase the accuracy of the identified transition probability models in dip and strike directions, as well as the simulated hydrofacies architectures.

**Acknowledgements**

This work was supported by the National Natural Science Foundation (No.41201420, 41130744), Beijing Nova Program (No.Z111106054511097) and Beijing Young Talent Program. We benefited from discussions with Robert W. Ritzi of the Wright State University and his comments and suggestions greatly improve this paper.

# Reference

Agterberg, F.P., 1974. *Geomathematics*, Elsevier Sci., New York.

Anderson, M.P., 2007. Introducing groundwater physics, *Physics Today*, 60(5): 42-47.

Carle, S.F. and Fogg G.E., 1996. Transition probability-based indicator geostatistics, *Mathematical Geology*, 28(4): 453-475.

Carle, S.F. and Fogg G.E., 1997. Modeling spatial variability with one and multimensional continuous-lag Markov chain, *Mathematical Geology*, 29(7): 891-918.

Carrera, J. and Neuman S. P., 1986. Estimation of aquifer parameters under steady state and transient condition: 2. Uniqueness, stability, and solution algorithms, *Water Resour. Res.,* 22(2), 211 – 227.

Clifton, P. M. and Neuman S. P., 1982. Effects of kriging and inverse modeling on conditional simulation of the Avra Valley aquifer in southern Arizona, *Water Resour. Res.,* 18(4): 1215-1234.

Dai, Z. and Samper, J., 2004. Inverse problem of multicomponent reactive chemical transport in porous media: Formulation and applications, *Water Resour. Res.*, 40, W07407.

Dai, Z., Ritzi, R.W., Dominic D.F., 2005. Improving permeability semivariograms with transition probability models of hierarchical sedimentary architecture derived from outcrop analog studies, *Water Resour. Res.*, 41, W07032.

Dai, Z., Wolfsberg, A., Lu Z., Reimus P., 2007a. Upscaling matrix diffusion coefficients for heterogeneous fractured rocks, *Geophys. Res. Lett.,* 34, L07408.

Dai, Z., Wolfsberg, A., Lu Z., Ritzi R. Jr, 2007b. Representing aquifer architecture in macrodispersivity models with an analytical solution of the transition probability matrix, *Geophys. Res. Lett., 34*, L20406.

Dai, Z., Wolfsberg, A., Lu Z. Deng H., 2009. Scale dependence of sorption coefficients for contaminant transport in saturated fractured rock, *Geophys. Res. Lett.*, 36: L01403.

Dai, Z., Wolfsberg, A., Reimus P., Deng H., Kwicklis E., Ding M., Ware D., Ye M., 2012. Identification of sorption processes and parameters for radionuclide transport in fractured rock, *J. of Hydrol.* 414-415: 220-230.

Dai, Z., Middleton, R., Viswanathan, H., Fessenden-Rahn, J., Bauman, J., Pawar, R., Lee, S. and McPherson, B., 2014. An integrated framework for optimizing $CO_2$ sequestration and enhanced oil recovery, *Environ. Sci. Technol. Lett.*, 1, 49-54.

Deng, H., Dai, Z., Wolfsberg, A. V., Lu, Z., Ye, M., Reimus, P., 2010. Upscaling of reactive mass transport in fractured rocks with multimodal reactive mineral facies, *Water Resour. Res*., 46: W06501.17

**List of Tables**

**Table 1. Statistical data of three alluvial fan zones observed from 783 boreholes**

| Zone | Borehole number | Area (Km$^2$) | Parameters | Sub-clay and clay | Fine sand | Medium-coarse sand | Gravel |
|---|---|---|---|---|---|---|---|
| Zone 1 | 52 | 159.85 | Proportion | 0.148 | 0.215 | 0.075 | 0.562 |
| | | | Mean thickness (m) | 7.63 | 9.08 | 8.78 | 17.3 |
| | | | Maximum cumulative thickness (m) | 99.7 | 147.8 | 95.8 | 336 |
| Zone 2 | 363 | 360.92 | Proportion | 0.423 | 0.220 | 0.087 | 0.270 |
| | | | Mean thickness (m) | 7.31 | 4.77 | 3.87 | 7.97 |
| | | | Maximum cumulative thickness (m) | 200 | 130.1 | 86.8 | 214 |
| Zone 3 | 368 | 633.81 | Proportion | 0.502 | 0.293 | 0.121 | 0.084 |
| | | | Mean thickness (m) | 8.53 | 6.20 | 4.57 | 5.78 |
| | | | Maximum cumulative thickness (m) | 252.8 | 200.1 | 120.3 | 101.3 |
| Whole area | 783 | 1154.58 | Proportion | 0.447 | 0.262 | 0.106 | 0.185 |
| | | | Mean thickness (m) | 8.07 | 5.82 | 4.46 | 8.21 |

Table 2. The estimated statistical parameters for hydrofacies in three zones

| Zone | Parameters | Categories | Estimated Parameter | Confidence interval (95%) |
|---|---|---|---|---|
| Zone 1 | Integral scale (m) | Vertical | 17.05 | (11.08, 23.03) |
| | | Dip | 618 | by trial and error |
| | | Strike | 309 | by assumption |
| | Volumetric Proportion | Sub-clay and clay | 0.1657 | (0.1229, 0.2085) |
| | | Fine sand | 0.2346 | (0.1919, 0.2773) |
| | | Medium-coarse sand | 0.0669 | (0.0236, 0.1101) |
| | | Gravel | 0.5328 | (0.4889, 0.5768) |
| Zone 2 | Integral scale (m) | Vertical | 6.245 | (2.664, 9.826) |
| | | Dip | 360 | by trial and error |
| | | Strike | 180 | by assumption |
| | Volumetric Proportion | Sub-clay and clay | 0.4093 | (0.3717, 0.4469) |
| | | Fine sand | 0.2854 | (0.2479, 0.3229) |
| | | Medium-coarse sand | 0.0653 | (0.0277, 0.1030) |
| | | Gravel | 0.2400 | (0.2025, 0.2775) |
| Zone 3 | Integral scale (m) | Vertical | 5.348 | (3.740, 6.957) |
| | | Dip | 319 | by trial and error |
| | | Strike | 159.5 | by assumption |
| | Volumetric Proportion | Sub-clay and clay | 0.5028 | (0.4839, 0.5217) |
| | | Fine sand | 0.3277 | (0.3088, 0.3465) |
| | | Medium-coarse sand | 0.1066 | (0.0877, 0.1255) |
| | | Gravel | 0.0629 | (0.0441, 0.0818) |

Table 3. The estimated statistical parameters for hydrofacies with a single-zone transition probability model in the study area

| Parameters | Categories | Estimated Parameter | Confidence interval (95%) |
|---|---|---|---|
| Integral scale (m) | Vertical | 8.233 | (3.77, 12.69) |
| | Dip | 469 | by trial and error |
| | Strike | 234.5 | by assumption |
| Volumetric Proportion | Sub-clay and clay | 0.4223 | (0.3807, 0.4639) |
| | Fine sand | 0.3033 | (0.2619, 0.3447) |
| | Medium-coarse sand | 0.0886 | (0.0470, 0.1302) |
| | Gravel | 0.1858 | (0.1444, 0.2272) |

**List of figure captions**

Figure 1. Location of the study area, distribution of the boreholes, 35 borehole logs (5 logs in Zone 1, 15 logs in Zone 2, and 15 logs Zone 3)

Figure 2. Volumetric proportions of four hydrofacies and integral scales in three zones

Figure 3. Sample (circle symbol) and computed (solid line) transition probabilities in vertical (a) and dip (b) directions in Zone 1

Figure 4.Sample (circle symbol) and computed (solid line) transition probabilities in vertical (a) and dip (b) directions in Zone 2

Figure 5.Sample (circle symbol) and computed (solid line) transition probabilities in vertical (a) and dip (b) directions in Zone 3

Figure 6.The averaged single-zone sample (circle symbol) and computed (solid line) transition probabilities in vertical (a) and dip (b) directions with all of the borehole data

Figure 7. Simulated three dimensional hydrofacies structure and cross sections in the study area (with 15 times of vertical exaggeration)

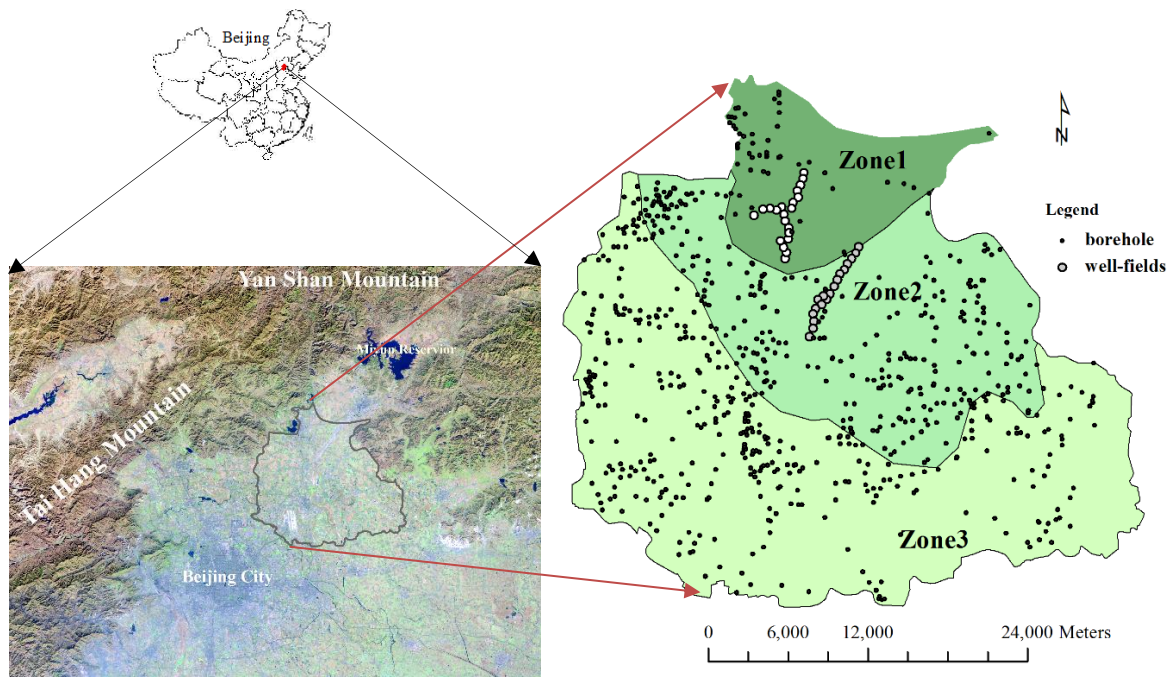

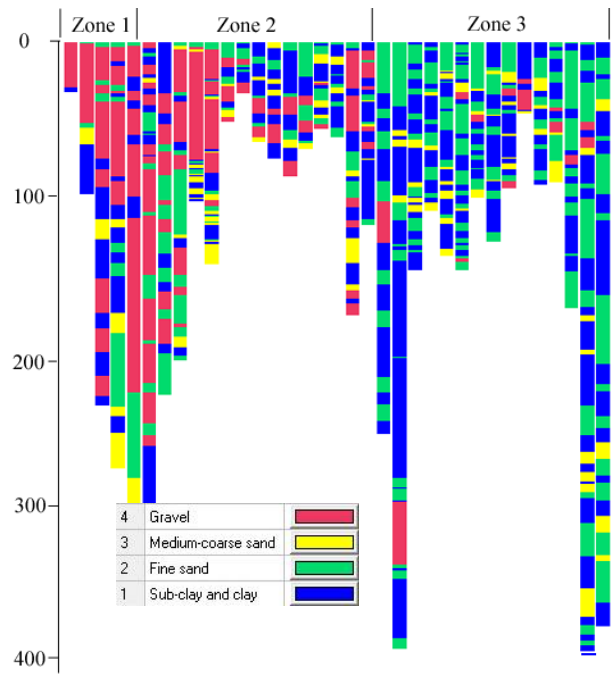

Figure 1

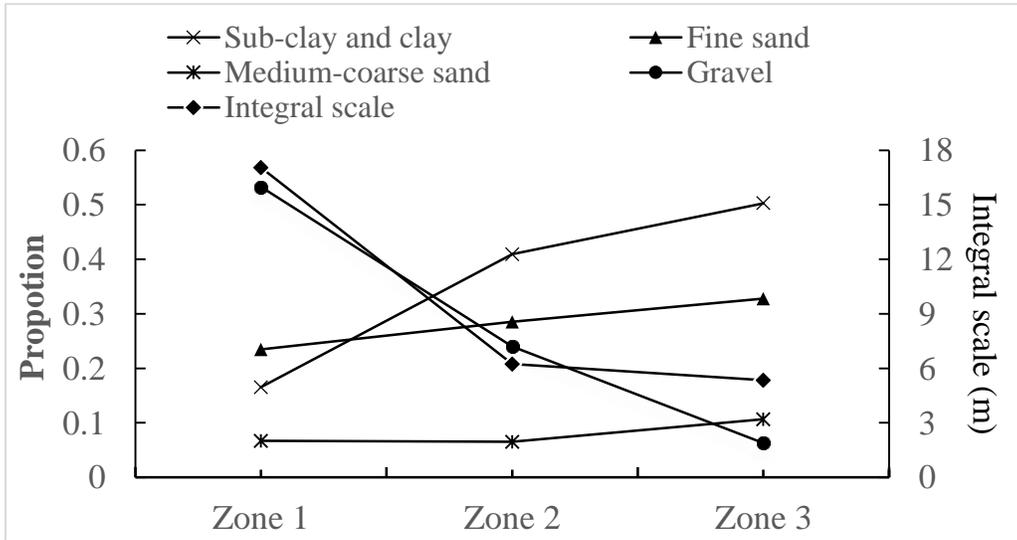

Figure 2

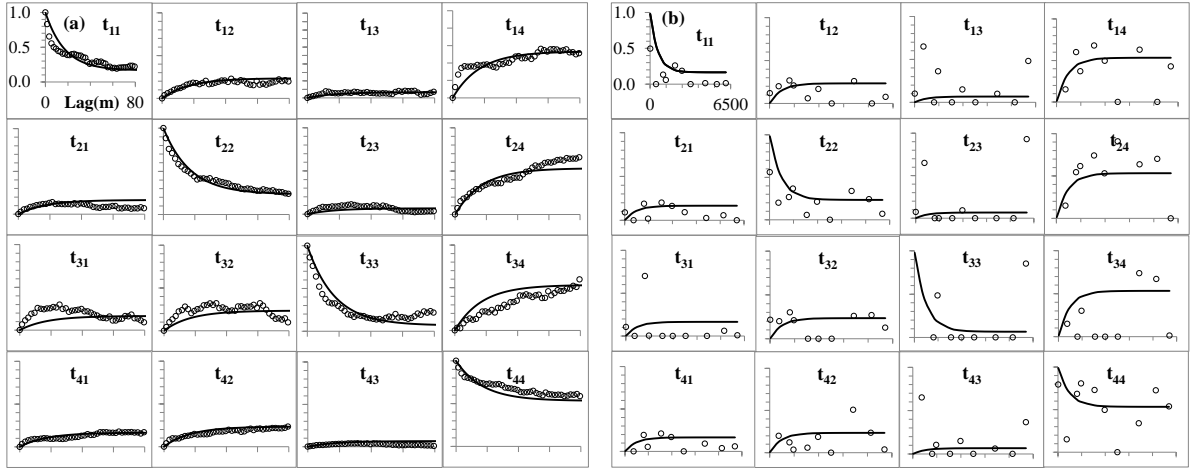

Figure 3

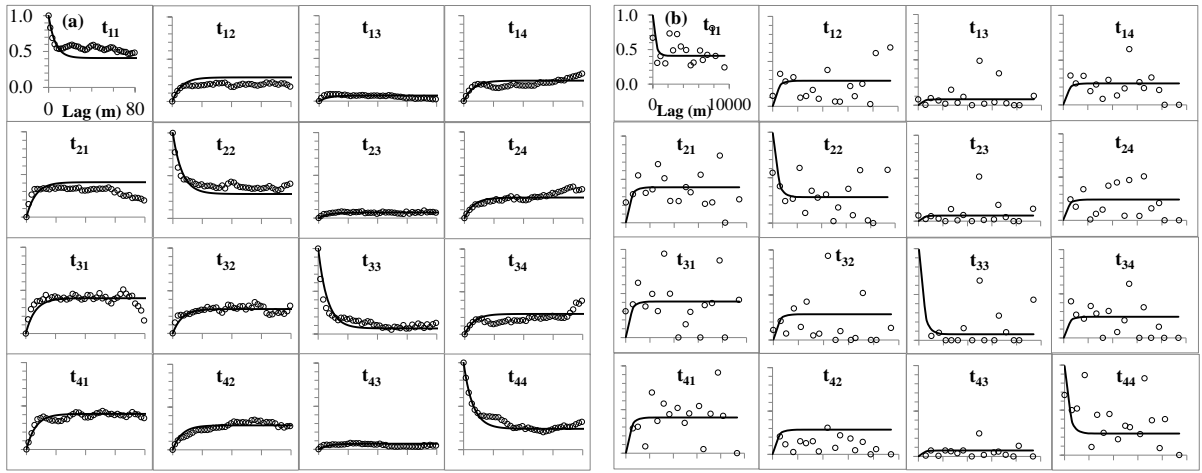

Figure 4

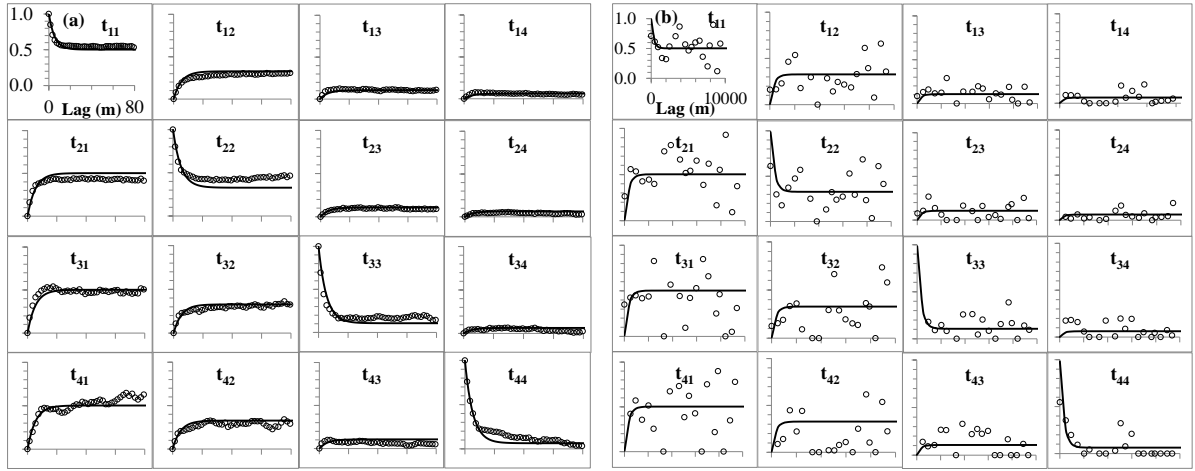

Figure 5

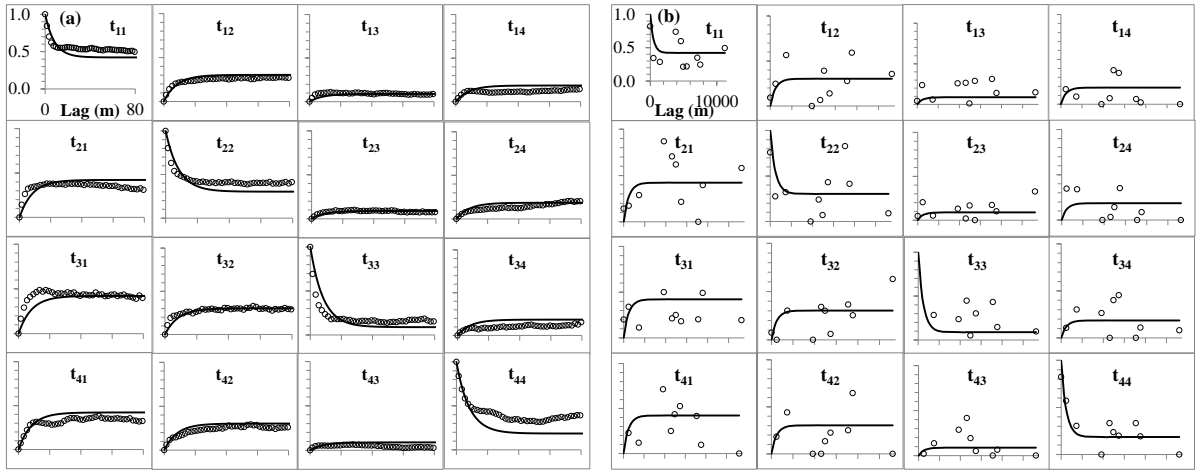

Figure 6

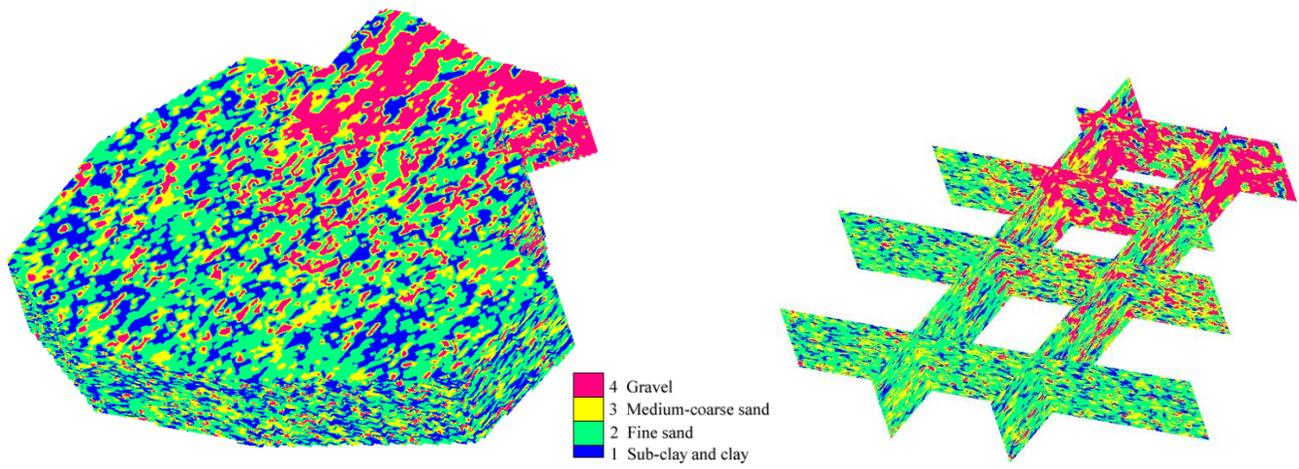

Figure 7